\documentclass[journal=jacsat,manuscript=article]{achemso}

\usepackage{amssymb,amsfonts,amsmath}
\usepackage{latexsym}
\usepackage{graphicx}
\usepackage{rotating}
\usepackage{caption}
\usepackage{tikz}
\usepackage[T1]{fontenc} 
\usepackage{soul}
\usepackage[normalem]{ulem}

\newcommand{\oxy}{O$_2$}
\newcommand{\butterfly}{Cu$_2$O$_2$}

\newcommand{\onlinecite}[1]{\nocite{#1}\citenum{#1}} 
\newcommand{\up}{\uparrow}
\newcommand{\dn}{\downarrow}

\graphicspath{{./images/}}

\usepackage{graphicx}
\makeatletter
\let\saved@includegraphics\includegraphics
\AtBeginDocument{\let\includegraphics\saved@includegraphics}
\renewenvironment*{figure}{\@float{figure}}{\end@float}
\makeatother

\author{Mohamed Ali al-Badri}
\affiliation[King's College London]{Theory and Simulation of Condensed Matter (TSCM), Strand, London WC2R 2LS, United Kingdom}
\email{Mohamed.al-Badri@kcl.ac.uk}
\author{Edward Linscott}
\affiliation{Cavendish Laboratory, University of Cambridge, J.\,J.\,Thomson Avenue, Cambridge CB3 0HE, United Kingdom}
\author{Antoine Georges}
\affiliation{Coll\`ege de France, 11 Place Marcelin Berthelot, 75005 Paris, France}
\alsoaffiliation{Center for Computational Quantum Physics, Flatiron Institute, 162 Fifth Avenue, New York, NY 10010, USA}
\alsoaffiliation{CPHT, Ecole Polytechnique, CNRS, Universit\'e Paris-Saclay, 91128 Palaiseau, France}
\alsoaffiliation{DQMP, Universit\'e de Gen\`eve, 24 Quai Ernest Ansermet, CH-1211 Gen\`eve, Suisse}
\author{Daniel J. Cole}
\affiliation{School of Natural and Environmental Sciences, Newcastle University, Newcastle upon Tyne NE1 7RU, United Kingdom}
\author{C\'edric Weber}
\affiliation[King's College London]{Theory and Simulation of Condensed Matter (TSCM), Strand, London WC2R 2LS, United Kingdom}
\email{Cedric.Weber@kcl.ac.uk}

\title{Superexchange mechanism and quantum many body excitations in the archetypal di-Cu oxo-bridge}

\begin{document}

\maketitle

\newpage

\begin{abstract}
We perform first-principles quantum mechanical studies of dioxygen ligand binding to the hemocyanin protein.
Electronic correlation effects in the functional site of hemocyanin are investigated using a state-of-the-art approach, treating the localised copper 3\emph{d} electrons with cluster dynamical mean field theory (DMFT) for the first time. 
This approach has enabled us to account for dynamical and multi-reference quantum mechanics, capturing valence and spin fluctuations of the 3\emph{d} electrons. Our approach explains the stabilisation of the experimentally observed di-Cu singlet for the butterflied \butterfly\ core, with localised charge and
incoherent scattering processes across the oxo-bridge that prevent long-lived charge excitations, suggesting that the magnetic structure
of hemocyanin is largely influenced by the many-body corrections. Our computational model is supported by agreement with experimental optical absorption data,
and provides a revised understanding of the bonding of the peroxide to the di-Cu system \emph{in vivo}.
\end{abstract}

\section{Introduction}

Copper-based metalloproteins play a major role in biology
as electron or dioxygen (\oxy) transporters.
Hemocyanin (Hc) is one of three oxygen transporting proteins found in
nature, alongside the iron-based hemerythrin and hemoglobin, and is
common to a number of invertebrates, such as molluscs and arthropods.
Deoxy-Hc employs two half-spin copper (I) cations, each coordinated by
the imidazole rings of three histidine residues, to reversibly bind
\oxy. Furthermore, various type 3 copper-based systems possess catalytic properties. Hc can decompose hydrogen peroxide into water and oxygen\cite{Ghiretti1956} and synthetic analogues have been shown to reversibly cleave the dioxygen bond\cite{Halfen1996} --- a mechanism that enables tyrosinase and catechol oxidases to oxidise phenols.\cite{Duckworth1970}

An accurate understanding of the electronic structure (spin and
charge) of the \butterfly\ core is essential to clarifying the
operation of dioxygen transport and would advance the design of synthetic catalysts that
employ dioxygen as a terminal oxidant. There is significant interest in the biomimetic application of naturally 
occurring metal complexes for use in metallodrug design, with Cu(II) 
complexes recently employed in cancer therapeutics as artificial DNA metallonucleases\cite{MCGIVERN} and tyrosinase mimics.\cite{NUNES}

However, the formation of oxygenated hemocyanin (oxyHc)
via the binding of \oxy\ to deoxygenated hemocyanin (Hc) remains a challenging problem. In particular, the binding of \oxy\ falls into the category of a spin-forbidden transition. 
Molecular \oxy\ is in a spin triplet configuration, and the Cu ions in deoxyHc are known to be in the Cu(I) $d^{10}$ singlet configuration. The combination of triplet \oxy\ and singlet deoxyHc, to produce the \butterfly\
antiferromagnetic singlet in oxyHc, is believed to occur via a simultaneous
charge transfer of one electron from each Cu(I) ion to \oxy\, forming a hybrid
Cu(II)-peroxy-Cu(II) configuration. A superexchange pathway 
is hypothesised to form across the two Cu atoms.\cite{Gherman2009} This mechanism is supported by 
SQUID measurements that report a large superexchange coupling\cite{Dooley1978} between the two Cu centers, and a diamagnetic
ground state.\cite{Solomon2011}

Despite intensive theoretical studies on the nature of the side-on coordinated \butterfly\ core, theoretical analysis has so far proved to be challenging for many
electronic structure methods including \textit{ab initio} quantum chemistry, density functional theory (DFT) and
mixed quantum/classical (QM/MM) methods due to the complex simultaneous interplay of both the charge and spin, and to the
multi-reference character of oxyHc. 
In particular, DFT and hybrid-DFT do not predict the correct singlet ground state
due to the multi-reference nature of the ground state that is not accessible in DFT-based approaches.\cite{takano2001, Metz2001, Saito2014}
To overcome the limitation of DFT techniques, a spin-projection method (also called spin-mixing) is often applied, whereby the different spin-polarised ground states are calculated individually,\cite{Cohen2007} and the entangled singlet is
reconstructed by linear combination of the respective Slater determinants
(essentially a combination of the spin-broken symmetry state in the up-down, up-up,
down-down configurations to extract an effective singlet state). 
Although this construction yields insights into the energetics,
it does not allow for the study of excitations, and limits the scope of comparison with experimental
data, such as the optical absorption,\cite{Metz2001} that is a stringent test of any theory.
Furthermore, the spin-contamination present in spin-polarised hybrid DFT remains
an issue,\cite{Saito2014,Cramer2006a,Cramer2006b,
Gherman2009,Siegbahn2006} and typically the broken symmetry state wrongly becomes 
asymmetric in the oxyHc \butterfly\, core.\cite{Metz2001}
Finally, experiments have also alluded to the necessity of characterising the oxyHc ground state as a mixed valence state,\cite{loeb}
which cannot be accessed in the ground state DFT approach.

Meanwhile, multireference wave function methods have been used to extensively study the oxyHc core,
\cite{Cramer2006a,Cramer2006b,Roemelt2016,Flock1999,Rode2005,Malmqvist2008,kurashige2014,Kurashige2009} 
but these approaches are not feasible for systems containing more than several dozen electrons. Flock and Pierloot\cite{Flock1999} argue that the inclusion of the imidazole ligands results in steric effects that are critical for a realistic description of oxygen containing dicopper systems, but most multireference wave function studies of this core consider a simplified model with ammonia ligands (including CC,\cite{Cramer2006a, Cramer2006b} CASPT2,\cite{Flock1999} MRCI,\cite{Rode2005} RASPT2,\cite{Malmqvist2008} and DMRG-CASPT2\cite{kurashige2014}), while others (such as DMRG\cite{Kurashige2009} and DMRG-CT\cite{yanai2010dmrgct}) are limited to the experimentally inaccessible bare \butterfly$^{2+}$ core. 
Furthermore, this system has large active space requirements given that it likely suffers from triplet instability,\cite{Saito2014} and if the number of allowed excitations is too limited, size-extensivity errors arise.\cite{Malmqvist2008}
Some of these methods also lack dynamical correlation contributions,\cite{malmqvist1989,Roos1987, Roos1980}
and others strongly over-correct correlation effects.\cite{Rode2005}


In this work, we apply an alternative strategy, namely dynamical mean-field theory (DMFT).\cite{Georges1996a} This method belongs to a broad category of embedding approaches and accounts for the limitations discussed above by
treating the many-body effects and the superexchange of the di-Cu bridge explicitly (unlike DFT), while limiting this treatment to the correlated subspace of the copper $3d$ electrons, thereby side-stepping the prohibitive scaling of quantum chemistry methods. DMFT
is a sophisticated method that includes quantum dynamical effects, and takes into account both valence and spin fluctuations, and thermal excitations. 
Although DMFT is routinely used to describe materials, it was also recently extended to molecular systems,\cite{referee1,referee2}
and combined with the linear-scaling DFT software, ONETEP, to extend the applicability of DFT\,+\,DMFT to systems of unprecedented size.\cite{onetep,our_paper_vo2,Linscott2018b,Cedric,Weber2014a} 

\section{Results}

\begin{figure}[t]
\begin{center}
\includegraphics[width=0.4\columnwidth]{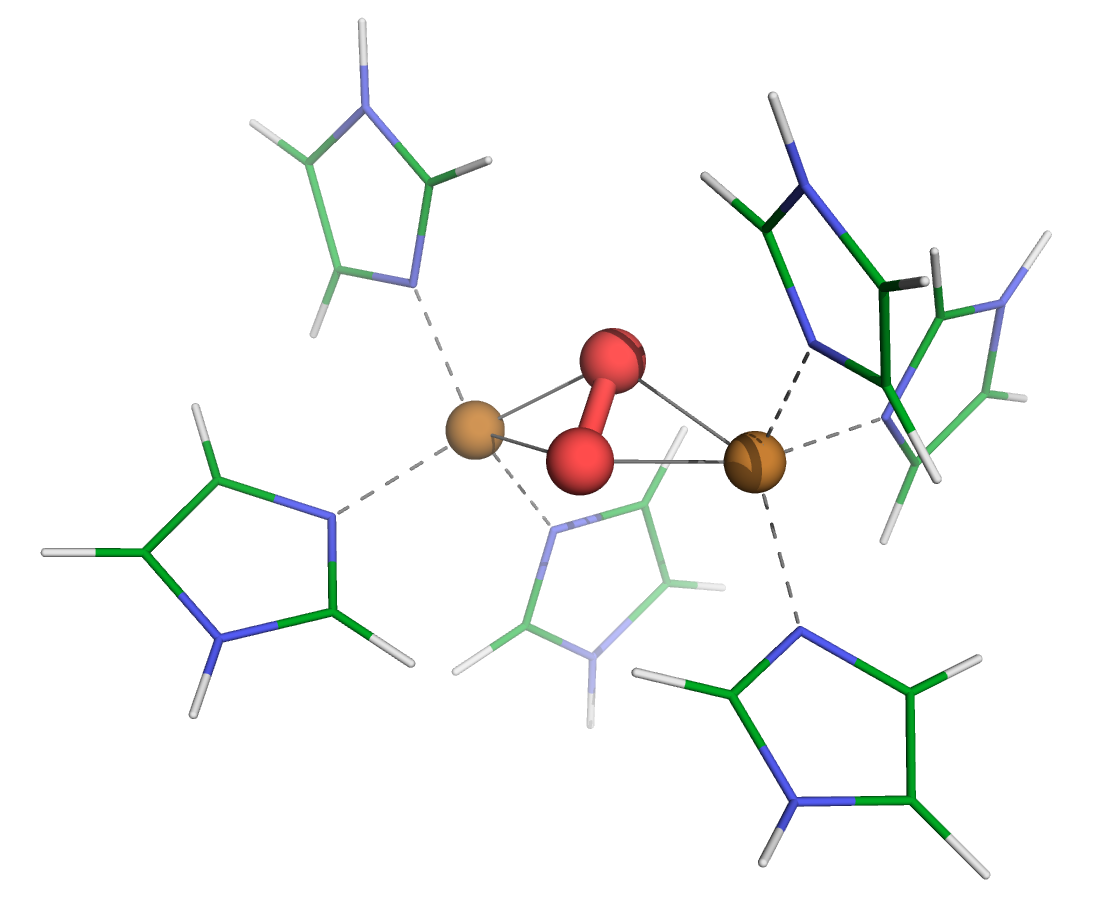}
\caption{The oxyHc functional complex, showing the \butterfly\  correlated subsystem, which is treated using DMFT, and the surrounding imidazole rings representing the protein environment.}
\label{fig_core}
\end{center}
\end{figure}

We perform the first DFT\,+\,DMFT simulations
on a 58-atom model of the oxyHc functional complex (Fig.~\ref{fig_core}).
Although previous DFT-based approaches have reported that a di-Cu singlet is obtained across the oxo-brige, these approaches
rely on artificial and approximate constructions to overcome the single-Slater determinant approach of DFT. And while 58 atoms is within the reach of some less accurate quantum chemistry methods, the overhead of extending DFT\,+\,DMFT to include much more of the protein environment would come at an insignificant computational cost.
Instead, we treat the multi-reference, finite-temperature and
explicit on-site Coulomb interaction effects associated with the Cu $3d$ binding site, in single, self-consistent calculations, and
systematically investigate how the Hubbard Coulomb potential $U$
alters the electronic structure at the \butterfly\ site.

We report the quantum entangled low energy states and analyse the dominant contributions
to the charge and magnetic properties of \butterfly. This enables the identification of a regime of parameters where the singlet is stabilised for the butterflied structure, in line with experiment ($U=6$ to $8$\,eV).  
We show that a dominant spin singlet state is maximised at $U = 8$\,eV and discuss the validation of the model by optical experiments. The obtained singlet is in the Heitler-London regime (entangled quantum superposition of two localised magnetic moments), and is associated with incoherent scattering processes that reduce the lifetime of charge excitations. 


As this problem involves direct exchange across two correlated atoms,
we use the non-local DMFT implementation (cluster DMFT)
needed to characterise the superexchange mechanism between the Cu$_2$\ \emph{d}-orbitals and intermediate \emph{p}-orbitals, as single site DMFT can only treat the multiplet structure of each Cu atom separately. 
In its simplest form, DMFT invokes a mean-field approximation across the multiple correlated sites, which
is not the case in our approach, as all correlated sites are directly treated in a multi-site Anderson impurity model (AIM) --- hence our approach might be denoted as DFT+AIM.
We do, however, carry out a self-consistency cycle over the charge density, as we work at fixed total number of electrons, and this produces a feedback to the solution of the model that is similar in its nature to the mean-field approximation used in DMFT.

The Hubbard $U$ correction is known to be paramount to describe many-body effects
responsible for the superexchange process across Cu atoms. Several competing effects stem
from the local Hubbard $U$ physics: charge localisation, exchange of electrons, charge-transfer excitations, and stabilisation of magnetic multiplets.  
Although typical values for $U$ can be obtained by linear response or constrained RPA,\cite{Scherlis2007a, LinscottPRB2018} these predictions are typically dependent on the choice of local orbitals and basis representation. 
We instead consider a range of values for $U$ (from 0 to 10\,eV). By artificially manipulating the magnitude of the local many-body effects, we can investigate their influence on the electronic spectral weight and magnetic properties.

\subsection{The spin state of the \butterfly\ core}
\begin{figure}[t]
\begin{center}
\includegraphics[width=0.7\columnwidth]{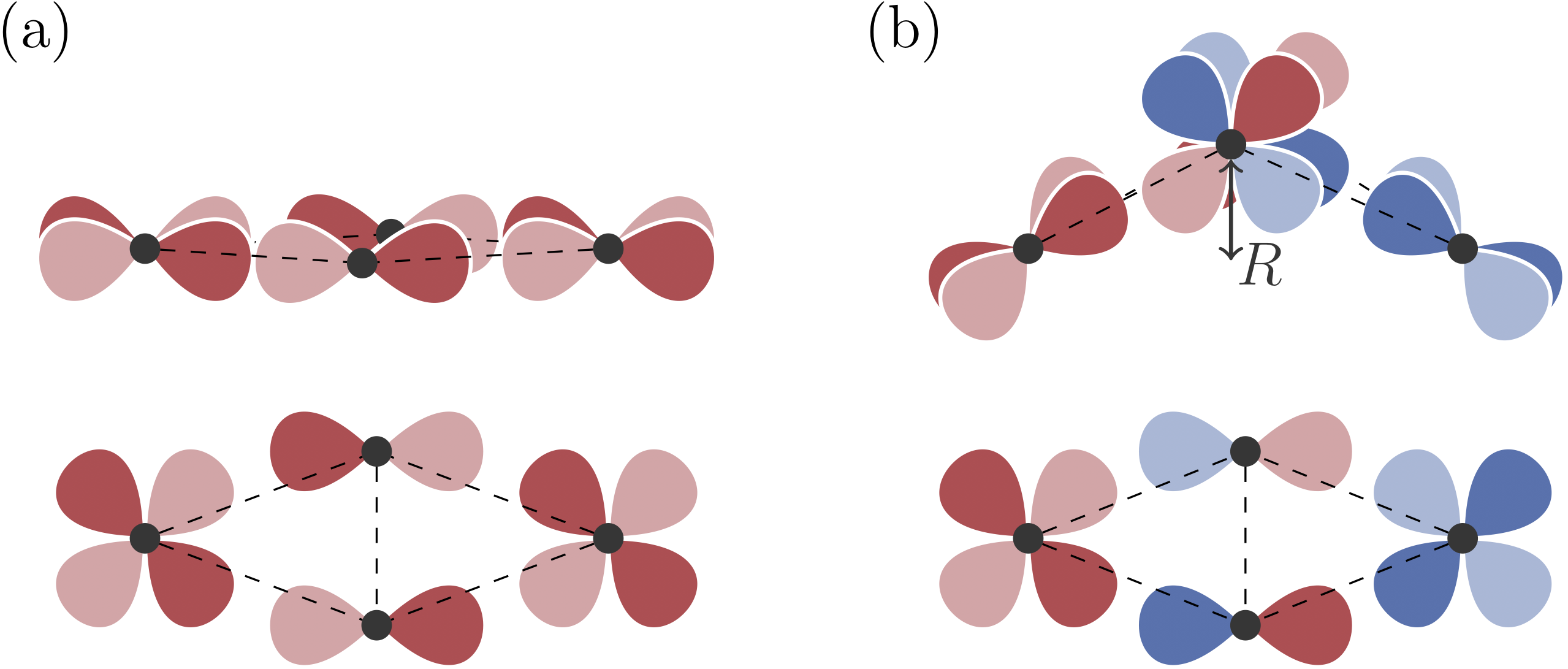}
\caption{The superexchange model of Solomon and co-workers, depicting the \butterfly\ core from side-on (top) and above (bottom). (a) In the planar configuration, single ligand orbitals bridge the two copper sites, and superexchange is possible. (b) In a bent configuration, the copper $d$ orbitals overlap with different $\pi^*$ orbitals. As these two sets of orbitals are orthogonal, hopping between the blue and the red subspaces is not possible and the superexchange mechanism breaks down.}
\label{fig_solomon_cartoon}
\end{center}
\end{figure}
\emph{In vivo}, the Cu$_2$O$_2$ core exists in a low-spin (singlet) state, as identified by EPR.\cite{Nakamura1960} 
In the model of Solomon and co-workers, this low-spin state is stabilised by superexchange via the O$_2$ ligand orbitals, which relies on the Cu$_2$O$_2$ core being planar (Fig.~\ref{fig_solomon_cartoon}(a)).\cite{Metz2001} As the peroxide molecule unbinds, the core butterflies (i.e. the dioxygen moves up out of the plane, leaving the core in a bent configuration). Here, each Cu overlaps with a different oxygen $\pi^*$ orbital on the peroxide (Fig.~\ref{fig_solomon_cartoon}(b)). This removes the superexchange, and the triplet state becomes most favourable. If we measure planarity by $R = |\frac{1}{2}(\mathbf{r}_\mathrm{CuA} + \mathbf{r}_\mathrm{CuB}) - \frac{1}{2}(\mathbf{r}_\mathrm{O1} + \mathbf{r}_\mathrm{O2})|$ --- that is, the distance between the mean position of the two copper atoms and the mean position of the two oxygen atoms --- the singlet-to-triplet transition occurs at $R = 0.6$\,\AA{} using the B3LYP DFT functional.\cite{Metz2001}

However, X-ray structures of the \butterfly\ core reveal that the bound singlet state is not planar. In oxyHc $R = 0.47$\,\AA, and in oxyTy $R = 0.63$\,\AA --- beyond the predicted singlet-to-triplet transition.\cite{Magnus1994,Matoba2006} QM/MM studies of the entire oxyHc protein (from which our model complex derives) obtain $R = 0.54$ to $0.71$\,\AA; evidently, the protein scaffolding around the binding site prevents the core ever reaching the planar structure observed in model complexes.\cite{Saito2014}

\begin{figure}[t]
\begin{center}
\includegraphics[width=0.4\columnwidth]{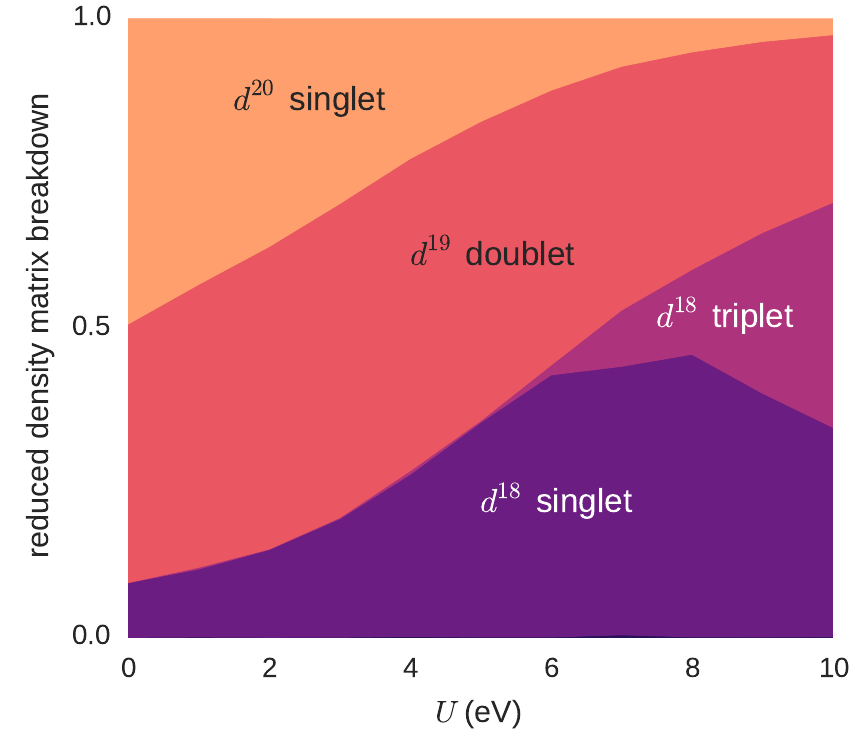}
\caption{Decomposition of the reduced density matrix of the Cu\textsubscript{2} dimer in the different quantum sectors. The colours correspond to the 
respective weights of the different contributions for each value of the Coulomb repulsion $U$ (if a colour occupies
all the vertical axis, for example, it means that all eigenvectors of the density matrix are in that particular quantum sector). 
Note that the $d$ occupation is the sum of both Cu sites (for example, $d^{20}$ means both Cu atoms are in the $d^{10}$ configuration).
}
\label{fig_spinfluc}
\end{center}
\end{figure}

With this in mind, we examined the reduced density matrix of our butterflied model ($R = 0.68$\,\AA; Fig.~\ref{fig_spinfluc}). This provides a detailed picture of the electronic structure of the $3d$ subspace
of the Cu atoms. Within DMFT, the density matrix is obtained by tracing the Anderson impurity model 
over the bath degrees of freedom, and gives an effective 
representation of the quantum states of the two Cu atoms. Note that in our approach,
the ground-state wave-function is not a pure state with a single 
allowed value for the spin states (singlet, doublet, triplet, 
\emph{etc.}), yet we can describe the fluctuating spin states of the 
Cu atom by analysing the spin distribution 
obtained from the dominant configurations. In the weakly correlated regime ($U<2$\,eV),
we obtain a large contribution from the $d^{20}$ and $d^{19}$ configurations, indicating that
the average charge transfer from the Cu to \oxy\ involves less than one electron per Cu,
thus preventing the formation of a singlet (as the Cu $3d$ orbitals are nearly full). 

As $U$ increases, the total electronic occupation of the Cu dimer decreases (Fig.~\ref{fig_spinfluc} and Table~S1). In the range $U=6-8$\,eV, the $d^{18}$ singlet component is maximised, beyond which $d^{18}$ triplet excitations begin to contribute. The existence of this singlet is corroborated by the observed local magnetic moment and spin correlation (Fig.~S1). Interestingly, the von Neumann entropy (Fig.~S2) grows as $U$ increases, pointing to the importance of many low-lying quantum states. Noting that $U \approx 8-8.5$\,eV in the case of both molecules\cite{10.1021/ja00209a041} and solids,\cite{Anisimov1991a} it appears that in nature many-body quantum effects stabilise the low-spin singlet in spite of the butterflied structure of the \butterfly\ core (Fig.~\ref{fig_solomon_cartoon}(b)).


\subsection{The superexchange mechanism}
Having identified this singlet in the \butterfly\ core at $U \approx 8$\,eV, let us establish how it forms. Direct hopping between localised Cu $d$-orbitals is very 
unlikely due to the large distance by which they are separated, 
and therefore hopping must proceed via an intermediate oxygen $p$-orbital (i.e. superexchange).\cite{Koch}
\begin{figure}
\begin{center}
\includegraphics[width=0.8\columnwidth]{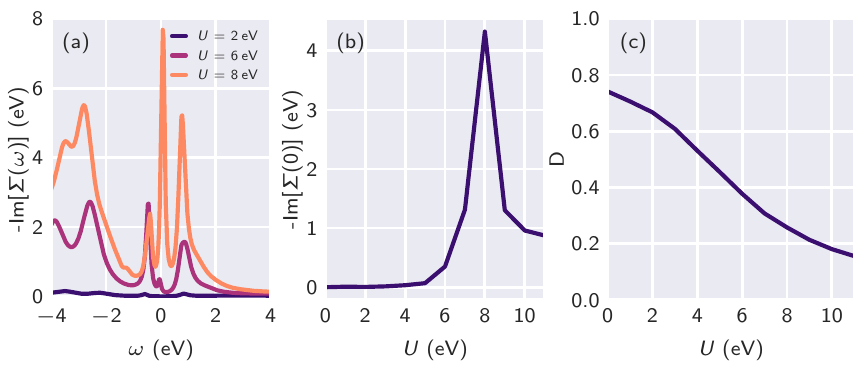}
\caption{
a) Imaginary part of the dynamical mean field local self energy of the Cu-$3d$ empty orbital for Hubbard $U=2$\,eV, $6$\,eV, and $8$\,eV.
At $U=8$\,eV, we obtain incoherent excitations at $\omega=0$\,eV. b) Self energy at $\omega=0$ and c) double occupancy $D$ as a function of $U$. Note that although the double occupancy is evolving smoothly with the Coulomb interaction $U$, $\Sigma(\omega=0)$ shows a sharp increase near $U=8$, associated with the stabilisation of a localised singlet.}
\label{fig_sigma}
\end{center}
\end{figure}

\begin{figure}
\begin{center}
\begin{tabular}{cc}
    \includegraphics[width=0.25\textwidth]{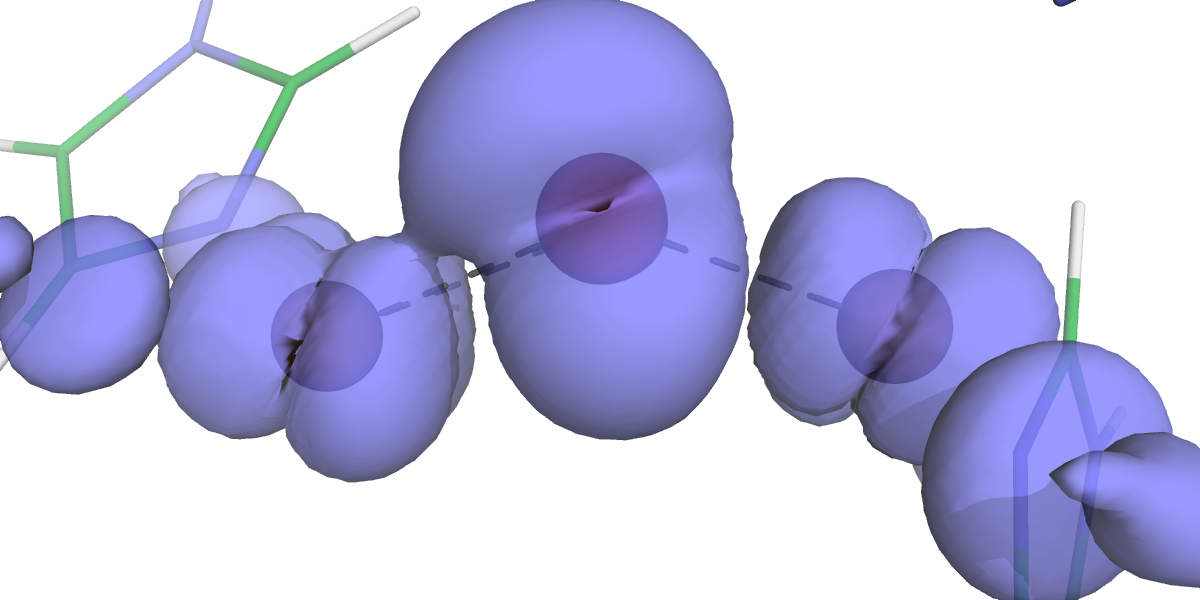} &
    \includegraphics[width=0.25\textwidth]{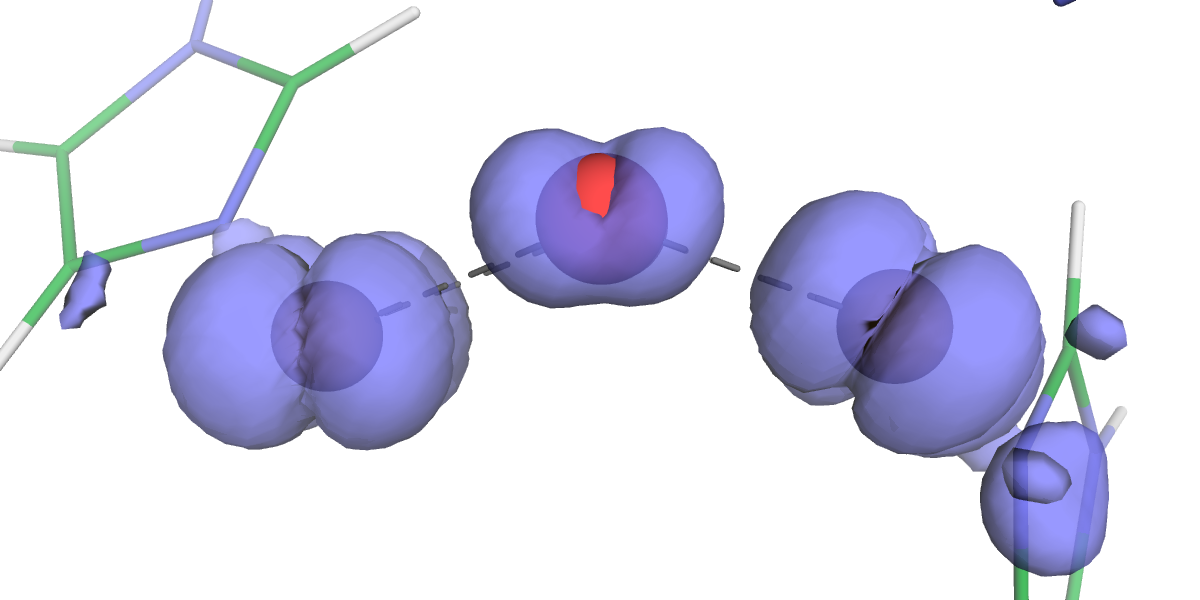} \\
    \includegraphics[width=0.25\textwidth]{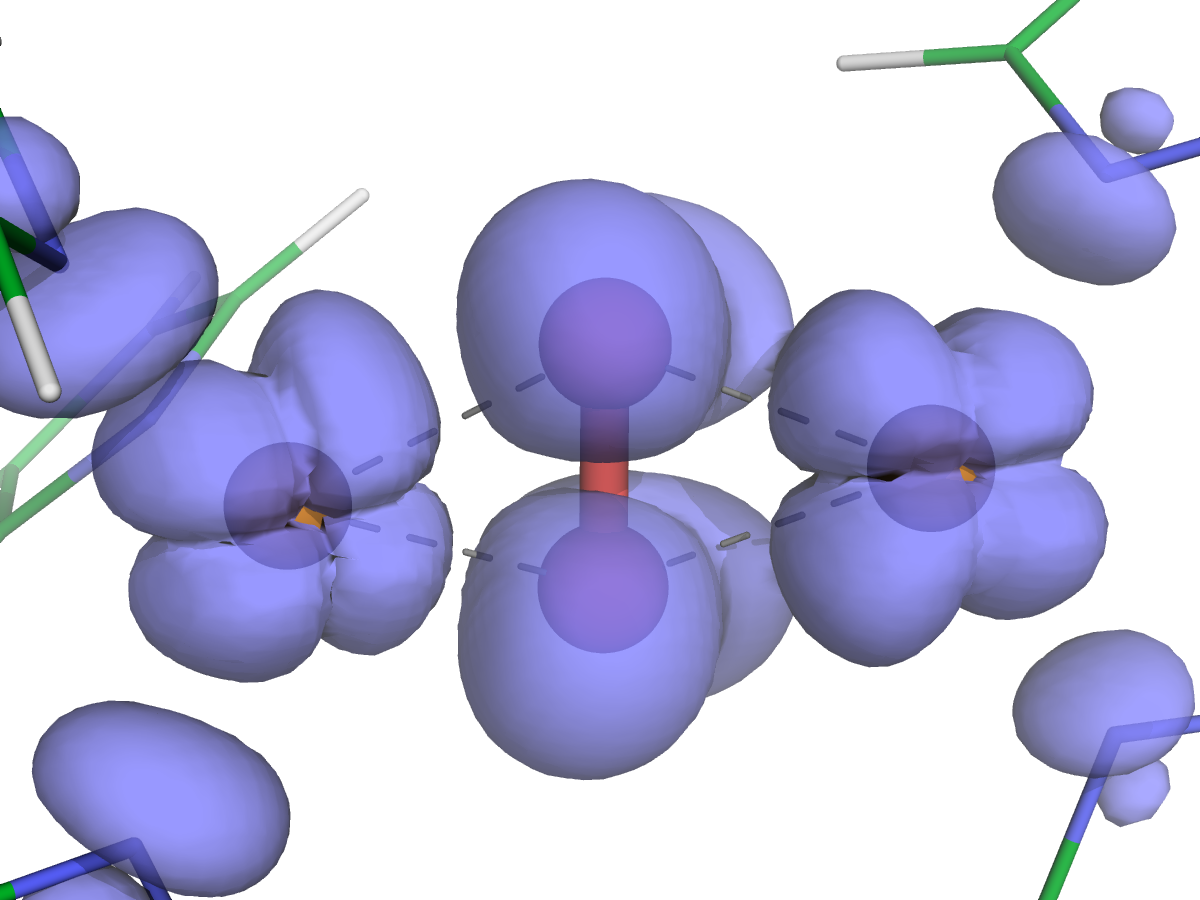} &
    \includegraphics[width=0.25\textwidth]{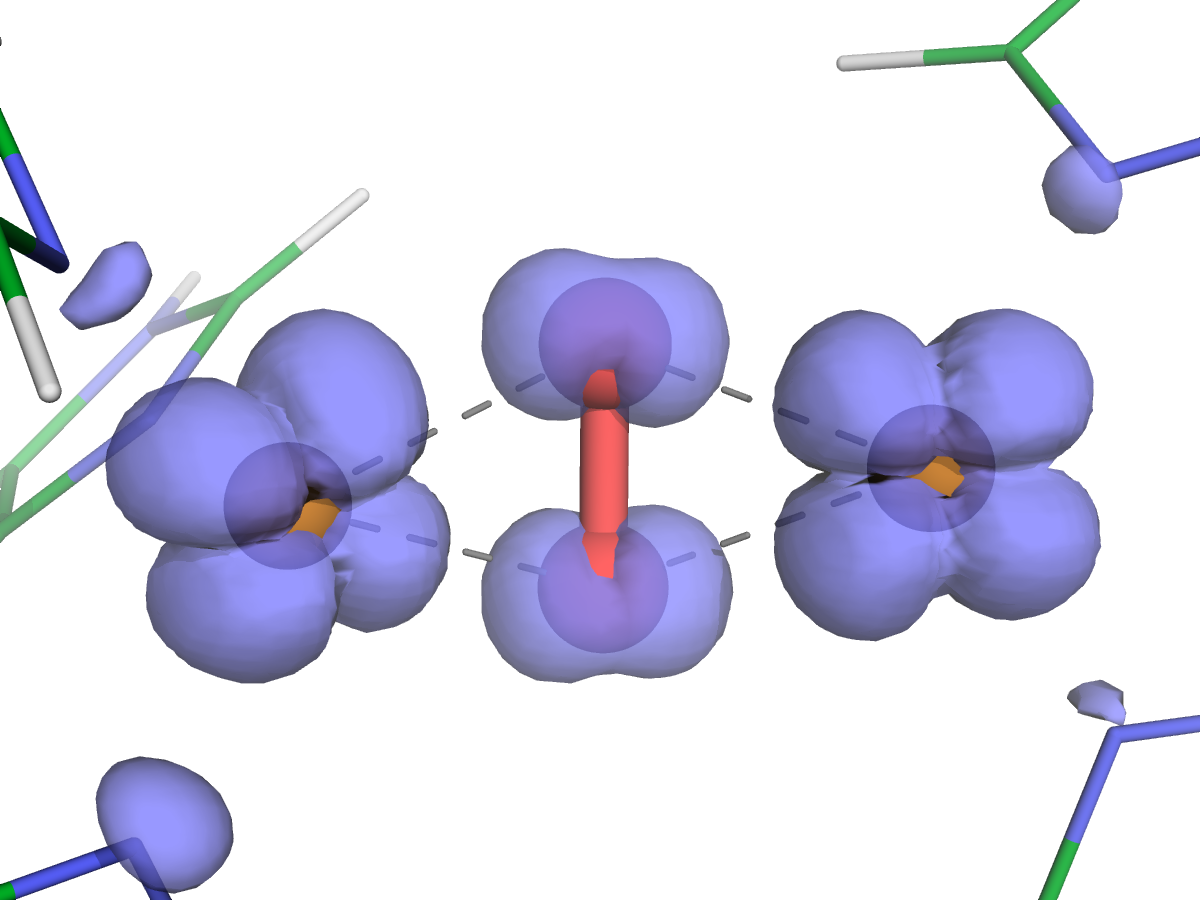} \\
\end{tabular}
\caption{Isosurfaces for the HOMO (left) and LUMO (right) densities for $U = 8$\,eV, as viewed side-on (top) and above (bottom). Note that because these are extracted from the Green's function via the spectral density, the phase of the orbitals is inaccessible.}
\label{fig_molecular_orbitals}
\end{center}
\end{figure}

The superexchange process can be pictured in the canonical hydrogen atom dimer system. This  system describes a pair of up and down electrons that form a singlet state.
In this picture, two different limits are possible: i) when the H atoms form a bond at short distance,
and the up and down electron form a delocalised bound singlet (BS) centred on the bond, with a high degree of double
occupancy, ii) in the dissociated case, known as the Heitler-London (HL) limit, where the H atoms are far apart and
the singlet is a true quantum entangled state of the singly occupied H orbitals. In the latter case, the molecular
orbital contains only one electron and double occupancy is zero. 
In the HL limit, which typically occurs in the limit of dissociation, the charge is localised
around the H atoms. However, this effect may also
occur in systems where the local Hubbard Coulomb repulsion $U$ acts as a Coulomb blockade: many-body effects prevent long-lived charge transfer excitations, and the Coulomb repulsion energy is reduced at the expense of the kinetic energy. A signature of the blockade is typically
a large increase in the self energy at the Fermi level (pole structure), indicating charge localisation and incoherent scattering associated with a short lifetime of charge excitations.

To investigate the nature of the singlet (BS or HL), we report in Fig.~\ref{fig_sigma}(a) the computed self energy of the Cu $3d$ subspace, for various values of $U$. We obtain a qualitative difference between $U=6$\,eV and $U=8$\,eV, where at $U=8$\,eV the self energy develops a pole at $\omega=0$. The formation of the pole is particular to $U=8$\,eV (Fig.~\ref{fig_sigma}(b)), and is associated with the regime where excitations are incoherent, which prevents long-lived charge transfer excitations from the Cu $3d$ orbitals to \oxy. In this limit, the many body effect acts as a Coulomb blockade and the charge is in turn localised, with weak direct coupling (HL limit). For $U=6$\,eV, the singlet is in the BS limit, where charge excitations allow a direct electron transfer across the oxo-bridge.  Note that the observation of the BS-HL crossover is not apparent in averaged quantities, such as in the double occupancies (Fig.~\ref{fig_sigma}(c)), which evolve smoothly with the Coulomb repulsion.

Turning back to the physical system, superexchange relies on a single ligand orbital linking the two copper sites. (Superexchange pathways via two $p$ orbitals are possible, but they give rise to ferromagnetic coupling that is significantly weaker than antiferromagnetic coupling from single-ligand orbital pathways\cite{Anderson1963}). Examination of the molecular orbitals near the Fermi level (Fig.~\ref{fig_molecular_orbitals}) reveals that, for the HOMO of the bent \butterfly\ structure, electronic density of the oxygen ligand is directed into the copper plane, thus providing a pathway for the antiferromagnetic superexchange that we observe.

Interestingly, we note that these molecular orbitals differ in their energetic ordering compared to those from DFT studies of planar model complexes.\cite{Solomon2011} In particular, the HOMO involves hybridisation with oxygen $\pi^*$ rather than $\sigma^*$ orbitals (with the $\sigma^*$ oxygen orbital featuring approximately 3\,eV above the Fermi level). This reordering will have substantial ramifications for the potential catalytic pathways, especially considering the importance of the $\sigma^*$ orbital to bond-breaking. This picture is confirmed by natural bond orbital (NBO) analysis, the results of which are presented in the Supplementary Information.

\subsection{Optical transitions} 
\begin{figure}[t]
\begin{center}
\includegraphics[width=0.8\columnwidth]{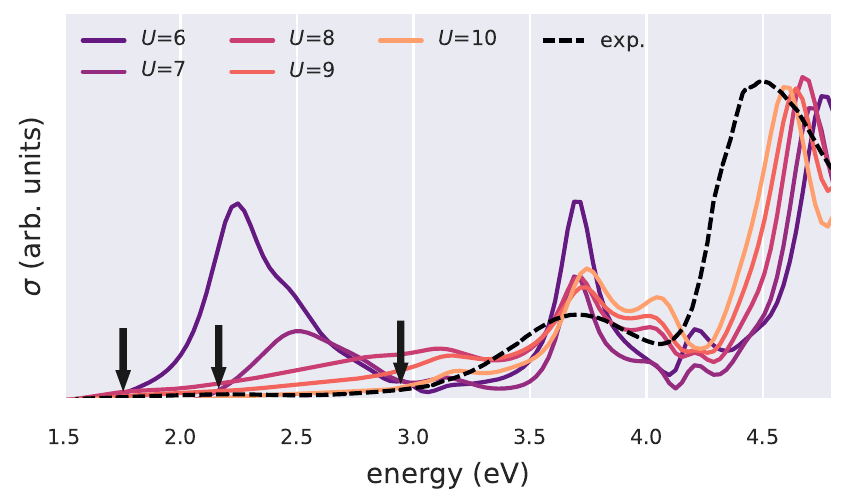}
\caption{
Theoretical optical absorption of the \butterfly\ core and imidazole rings obtained by DMFT for values of the Coulomb repulsion $U=6$ to $10$\,eV. For comparison, we show the experimental optical absorption\cite{andersen2011} in a wide range of wavelengths (infrared to UV). There are several smaller peaks in the experimental spectra that are not visible at this scale (indicated with arrows).}
\label{fig_optics}
\end{center}
\end{figure}

As a validation of the DFT\,+\,DMFT computational model, and to identify the strength of correlations in oxyHc, we extracted the optical absorption spectrum of ligated hemocyanin (Fig.~\ref{fig_optics}). 

As experiments are performed in the gaseous phase and not in a single crystal, we have calculated the isotropic and anisotropic components of the dielectric tensor. The former involves only pair correlators along the same spatial directions, whereas the latter also incorporates non-diagonal terms (note that both are causal quantities), which are important in oxyHc as the local coordination axes of the Cu atoms are not aligned. Remarkable agreement is obtained for $U=8$\,eV, with however a blue shift at high frequency ($\omega>4$\,eV), and a concomitant transfer of optical weight to lower energies ($\omega<3$\,eV). The position of the peak at approximately $3.5$\,eV is in excellent agreement. The experimental features at $3$\,eV (see arrow) are also visible in the theoretical calculations. The peaks at $1.8$\,eV and $2.2$\,eV contribute to the long infrared tail of the optical weight down to $1.5$\,eV. 
We note that the main difference between the BS ($U\leq 6$\,eV) and the HL singlet ($U=8$\,eV) is the presence of several large peaks associated with charge transfer from ligand to Cu $3d$ orbitals
below $2.5$\,eV, which are significantly smaller/absent in experiment.\cite{himmel,heirwegh} The suppression of the optical weight at $2$\,eV is due to a large increase in incoherent
scattering at $\omega=0$\,eV at $U=8$\,eV (Fig.~\ref{fig_sigma}), associated with the localisation of the holes
in the Cu $3d$ shell at $U=8$\,eV. The strong suppression of the optical weight in the near infrared regime and the consistent agreement with our calculations shows that the di-Cu singlet is in the HL regime. In comparison, DFT, without extensions, puts a strong emphasis on the near infrared peak in the optical absorption\cite{Metz2001} because the aforementioned scattering processes are absent at this level of theory. (See also the very small HOMO-LUMO gap in Fig.~S4 for small values of $U$.)

The absorption spectrum of this protein has been reported experimentally to 
be qualitatively dependent on its ligation state. In its oxygenated form, a 
weak peak at 570\,nm (approximately 2.2\,eV, see arrow) is attributed to ligand-to-metal charge 
transfer\cite{himmel, heirwegh} from the \oxy\ $\pi$ anti-bond with lobes 
oriented perpendicular to the metal atoms. This orbital is denoted 
``$\pi^*_v$"; the $\pi$ anti-bond with lobes directed towards the copper 
atoms is denoted $\pi^*_\sigma$ and is responsible for an experimentally observed (and much larger) peak 
at approximately 3.6\,eV. Our calculations accurately reproduce this large peak, associating it with a transition from the weight at -1\,eV in the density of states (Fig.~S4), which is localised on  Cu\textsubscript{A}/Cu\textsubscript{B}/imidazole, to the LUMO, which in our calculations is a hybridised state between
the Cu and the $\pi^*_\sigma$ O$_2$, with dominant Cu $3d$ character (Fig.~\ref{fig_molecular_orbitals}b). Meanwhile, a small peak at 2\,eV exists corresponding to the HOMO-to-LUMO transition ($\pi^*_v$ to Cu charge transfer).
Overall, comparison with the experimental spectrum suggests $U\geq 8$\,eV which is compatible with the value of $U$ that gives rise to the singlet.
We note that time dependent density functional theory (TDDFT) calculations reproduce the absorption peak at 3.6\,eV, but not the peak at 4.5\,eV. More importantly, the ground state electronic structure at this level of theory is not the antiferromagnetic singlet state observed experimentally, thus agreement between TDDFT and experiment likely involves some cancellation of errors.


\section{Conclusions}

We have presented the application of a DFT\,+\,cluster DMFT\ approach,
designed to treat strong electronic interaction and multi-reference 
effects, to oxyHc, a molecule of important biological function.
%
The reduced density matrix of the $3d$ subspace of the two Cu atoms revealed the presence of fluctuating spin-states, in which a Cu$_{2}$ $d^{18}$ singlet component is maximised at $U=8$\,eV in spite of the butterfly distortion of the \butterfly\ core.
The Hubbard $U$ is necessary to capture the multi-reference character of the ground-state, placing oxyHc in the limit of a true quantum entangled singlet in the limit of the Heitler-London model, while the highest occupied molecular orbital has been shown to provide a pathway for antiferromagnetic superexchange.
This work provides a starting point for studying biological activity of oxyHc and related type 3 Cu-based enzymes by (a) establishing that the singlet can survive the butterfly distortion, thereby resolving a prior inconsistency between structural data, spectroscopy, and first-principles calculations, and (b) by providing a framework for subsequent studies to account for the effect of the protein ``scaffolding" in which the active site sits, as well as the effects of strong electronic correlation.
Our approach reproduced the experimentally-observed peaks in the absorption spectrum at around 2.2\,eV, 3.7\,eV, and 4.5\,eV. 

The DFT\,+\,DMFT method provides a useful middle-ground between (a)
density-functional theory based simulations of metalloproteins, which depend heavily on the choice of functional and incorrectly describe multi-reference effects, and (b) single- or multi-reference quantum chemistry approaches, whose unfavourable scaling prevents studies on models of realistic size.
More generally, our particular implementation of DMFT within linear-scaling DFT software \cite{Cole2016a} can account for strongly correlated electronic behaviour while simultaneously including the effects of protein environments, making it ideally suited for studying biological activity in a wide range of transition metal-containing proteins.\cite{enzyme_transition_metal,Suga2015a}

\section{Methods}

The geometry of the 58 atom system was obtained from the literature.\cite{Saito2014} This had been optimised using the B3LYP hybrid functional and 
closely matches the experimentally observed structure.\cite{Magnus1994,Saito2014} The coordinates of this structure are provided in the Supplementary Information.
 
The DFT ground-state was obtained using ONETEP,\cite{onetep,ORegan2012a,Hine2010a} which is a linear-scaling DFT code that is formally equivalent to a plane-wave method. Linear-scaling is achieved by the \emph{in situ} 
variational optimisation of its atom-centered basis set (spatially-truncated 
nonorthogonal generalised Wannier 
functions, or NGWFs).\cite{Skylaris2002a} The total energy is directly minimised with respect to the NGWFs and the single-particle density matrix.
The use of a minimal, optimised Wannier function 
representation of the density-matrix
allows for the DFT ground state to be solved 
with relative ease in large systems. This is particularly useful in molecules, since explicit truncation of the basis functions ensures that the addition of vacuum does not increase the computational cost.

The DFT calculations of the oxyHc system were run with an energy cut-off of 897\,eV, using the Perdew-Burke-Ernzerhof (PBE) exchange-correlation functional.\cite{Perdew1996a} Nine NGWFs were employed on the copper atoms, four on each carbon/nitrogen/oxygen, and one on each hydrogen. Spin symmetry was imposed. NGWFs were truncated using 7\,\AA\,cutoff radii. Open boundary conditions were achieved via a padded cell and a Coulomb cut-off.\cite{Hine2011b} The Hubbard projectors were constructed from the Kohn-Sham solutions to an isolated copper pseudopotential.\cite{Ruiz-Serrano2012a} The pseudopotentials were generated with the OPIUM pseudopotential generation project.\cite{opium} These pseudopotentials partially account for scalar relativistic effects. (Studies have demonstrated that relativistic effects can play a role in the electronic structure of the Cu$_2$O$_2$ core\cite{Liakos2011}.)

We refined our DFT calculations using the DFT\,+\,DMFT 
method\cite{Georges1996a,dca_reference_cluster_dmft}
in order to obtain a more accurate treatment of strong electronic correlation effects. 
The oxyHc model was mapped, within DMFT, to an 
Anderson impurity model (AIM) Hamiltonian,\cite{heme_aim_kondo}
and we used a recently developed extended Lanczos solver\cite{cpt_ed_solver} to obtain the DMFT self energy. The convergence of the mapping is shown in Fig.~S5.

To identify the best spatial representation of the local $d$-space in the projected Anderson impurity model, we first identified the orthogonal transformation which reduces the off-diagonal elements of the local Green's function, for respectively the Cu\textsubscript{A} and Cu\textsubscript{B} sites. Then, we implemented a minimisation procedure which finds the closest corresponding real space $SO(3)$ rotation of the local Cartesian axis corresponding to the $O(5)$ orthonormal transformation in $d$-space. This provides a set of local axes, different on each Cu atom, which make the local Green's function nearly diagonal in frequency space. These axes are plotted in Fig.~S6. As a result, we observed that the Cu\textsubscript{A} and Cu\textsubscript{B} holes are of pure orbital character within this local axis representation. The validity of this approach was vindicated by the NBO analysis, which when performed on the DMFT density revealed a hole in one $3d$ orbital for each Cu atom, confirming the expected Cu(II) oxidation state 3$d^{9}$4$s^{0}$. (See Supplementary Information for details.)

Since only a single impurity site ($3d$ orbital subspace) is present, the system becomes crystal momentum independent in the molecular limit, and since the Kohn-Sham Green's function is computed in full before projection onto the impurity subspace, the Anderson impurity mapping is effectively exact, and the necessity of invoking the DMFT self-consistency is not required. However, in DFT\,+\,DMFT there is also a charge self-consistency cycle, where (i) the chemical potential can be updated to ensure particle conservation and/or (ii) the DFT\,+\,DMFT density kernel can be used to generate a new Kohn-Sham Hamiltonian, which in turn provides a new input to DMFT; the procedure being repeated until convergence is achieved.\cite{Pourovskii2007,Park2014,Bhandary2016} In this work, we updated the chemical potential but not the Hamiltonian due to computational cost.


To obtain the Kohn-Sham Green's function, we performed the matrix inversion, as well as all 
matrix multiplications involved in the DMFT algorithm,
on graphical processing units (GPUs) using a tailor-made parallel implementation of the Cholesky decomposition 
written in the CUDA programming language. 
Electronic correlation effects are described within the localised subspace by the 
Slater-Kanamori form of the Anderson impurity Hamiltonian,\cite{slater_kanamori_interaction,Kanamori1959a} specifically:
\begin{align}
\label{hint}
\mathcal{H}_{U} ={}& U \sum \limits_{m} {n_{m \up} n_{m \dn}} + \left(
U' -\frac{J}{2} \right)\sum\limits_{m>m'}{n_{m}n_{m'}}  \\
\nonumber {}&
-J \sum\limits_{m>m'}{\left( 2 \mathbf{S}_m \mathbf{S}_{m'} + \left(  d^\dagger_{m\up} d^\dagger_{m \dn } d_{m' \up} d_{m' \dn}   \right) \right)    },
\end{align}
where $m, m'$ are orbital indices, $d_{m\sigma}$ ($d^\dagger_{m\sigma}$) annihilates (creates) an electron with spin $\sigma$ in the orbital $m$, and $n_m$ is the orbital occupation operator.

The first term describes the effect of 
intra-orbital Coulomb repulsion, parameterised by $U$, and the second
term describes the inter-orbital repulsion, proportional to $U' = U - 2J$,
which is renormalised by the Hund's exchange coupling 
parameter $J$ in order
to ensure a fully rotationally invariant Hamiltonian (for 
further information on this topic, we
refer the reader to Ref.~\onlinecite{Imada1998}). 
The third term is the Hund's rule exchange coupling, 
described by a spin exchange coupling of amplitude $J$.
$\mathbf{S}_{m}$ denotes the spin corresponding to orbital $m$,
so that $S_{m}=\frac{1}{2}d_{m s}^{\dagger}
\vec{\sigma}_{s s'}
d_{m s' }$, where $\vec{\sigma}$ is the vector of Pauli matrices.

Our DMFT calculations were carried out at room 
temperature, $T=293$\,K and the Hubbard $U$ was varied in the range 
0\,--\,10\,eV, with a fixed Hund's coupling $J=0.8$\,eV.
The theoretical optical absorption was obtained in DFT\,+\,DMFT 
within the linear-response regime (Kubo formalism),
in the no-vertex-corrections approximation,\cite{millis_optical_conductivity_review} 
which is given by:
\begin{align}
 \sigma(\omega) ={}&  \frac{2 \pi e^2 \hbar }{\Omega }  \int  
 d \omega' \;
 \frac{  f( \omega'-\omega)-f \left( \omega' \right)}{\omega} \\
 {}&
 \times 
 \left( \mathbf{\rho^{\alpha \beta}} \left( \omega' - \omega \right)  \mathbf{v}_{\beta \gamma}  \mathbf{\rho^{\gamma \delta}} \left(\omega' \right) \cdot 
 \mathbf{v}_{\delta \alpha} \right), \nonumber
\end{align}
where the factor of two accounts for spin-degeneracy, 
$\Omega$ is the simulation-cell volume, $e$ is the electron charge,
$ \hbar $ is the reduced Planck constant, 
$f \left( \omega \right)$ is the Fermi-Dirac distribution, 
and $\mathrm{\rho}^{\alpha \beta}$ is the  density-matrix
given by the frequency-integral of the interacting DFT\,+\,DMFT Green's 
function.
The matrix elements of the velocity operator, $\mathbf{v}_{\alpha \beta}$,
noting that we do not invoke the Peierls substitution,\cite{millis_optical_conductivity_review}
are given by:
\begin{equation}
\mathbf{v}_{\alpha \beta}= 
- \frac{i \hbar }{ m_e }
\langle \phi_\alpha \rvert  \mathbf{\nabla} \lvert \phi_\beta \rangle
+ \frac{i}{\hbar} 
\langle \phi_\alpha \rvert  
\left[ \hat{V}_{nl}, \mathbf{r} \right] \lvert \phi_\beta \rangle.
\end{equation} 
This expression is general to the 
NGWF representation used in this work,\cite{optical_conductivity_non_orthogonal_basis}
where the contribution to the non-interacting  
Hamiltonian due to the non-local part of the 
norm-conserving pseudopotentials,\cite{PhysRevB.84.165131,
PhysRevB.44.13071} represented by $\hat{V}_{nl}$, is included.

Finally, the double-counting correction $E_{DC}$ must be introduced, since the contribution of interactions between the correlated orbitals to the total energy is already partially included in the exchange-correlation potential derived from DFT.  
The most commonly used form of the double-counting term is\cite{hund_coupling_averaged_double_counting}:
\begin{equation}
E_{\mathrm{dc}} = \frac{U^{\textrm{av}}}{2} n_d \left( n_d - 1 \right)  - \frac{J}{2} \sum\limits_\sigma{n_{d\sigma} \left( n_{d\sigma}-1 \right)}.
\end{equation} 
Natural bond orbital analysis was performed using the \emph{NBO 5} programme.\cite{nbo5} Performing this transformation starting from ONETEP's basis of NGWFs is described in Ref.~\onlinecite{Lee2013a}.
\section*{Data availability}
The data underlying this publication are available on the Apollo - University of Cambridge Repository. [link pending]
\section*{Code availability}
The ONETEP code version 5 is available from www.onetep.org.

\begin{acknowledgement}
This work was supported by BBSRC (grant BB/M009513/1), EPSRC (grants EP/N02396X/1, EP/L015552/1) and the Rutherford Foundation Trust. The Flatiron Institute is a division of the Simons Foundation. C.W. gratefully acknowledges the support of NVIDIA Corporation with the donation of the Tesla K40 GPUs used for this research. For computational resources, we were supported by the ARCHER UK National Supercomputing Service and the UK Materials and Molecular Modelling Hub for computational resources (EPSRC Grant No. EP/P020194/1).
\end{acknowledgement}

\subsection{Author contributions}
 C.\,W.\ and M.\,A.\,B. conceived and planned the research. M.\,A.\,B., E.\,B.\,L.\ and C.\,W.\ performed the calculations. All the authors analyzed the data and contributed to the final manuscript.
\subsection{Competing Interests} The authors declare that they have no
competing financial interests.
%
\bibliography{article}

\end{document}


\newpage
\section*{Details of diamagnetism}
\begin{figure}[h!]
\begin{center}
\includegraphics[width=0.8\columnwidth]{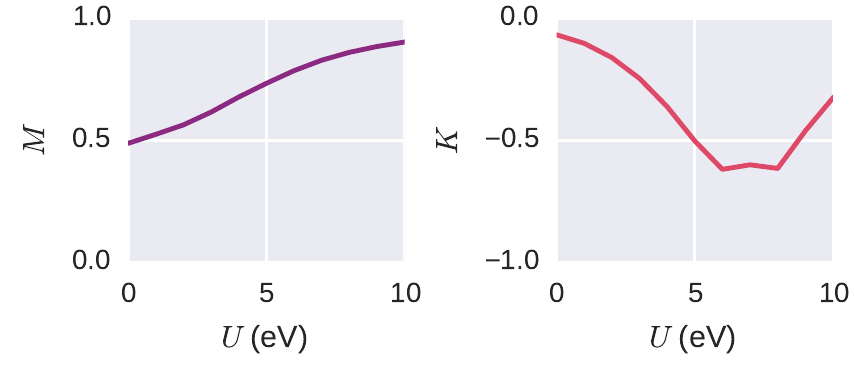}
\caption{The effective magnetic moment $M=\sqrt{\langle \mathbf{S}^2_1 \rangle/3}$ 
  (normalised by saturation value) and the spin correlation $K=2\langle \mathbf{S}_1 
  \cdot \mathbf{S}_2 \rangle$ for varying values of the Hubbard $U$. For a pure two orbital singlet, $K=-1.5$. In our calculations, as the rest of the molecule hybridises with the Cu orbitals, the spin-correlation is re-normalised to half its saturation value for $U=6-8$\,eV.}
\label{fig_spincorr}
\end{center}
\end{figure}

The effective magnetic moment and spin correlation of the Cu dimer are shown in Fig.~\ref{fig_spincorr}. The spin correlation $K$ reaches half the saturation
value for $U=6-8$\,eV. Note that the saturation value would only
be obtained for a diatomic system in vacuum, which is not hybridised to the
rest of the molecule. As the local Cu $3d$ orbital charge and spin
are not true quantum numbers in the molecule due to hybridisation, 
quantum fluctuations reduce the amplitude of the spin correlation to half the full
value. This is consistent with the presence of approximately $50\%$ combined $d^{19}$ and $d^{20}$ excitations (Fig.~3 of the main text).

\newpage
\section*{Von Neumann entropy} 
\begin{figure}[!ht]
\begin{center}
\includegraphics[width=0.4\columnwidth]{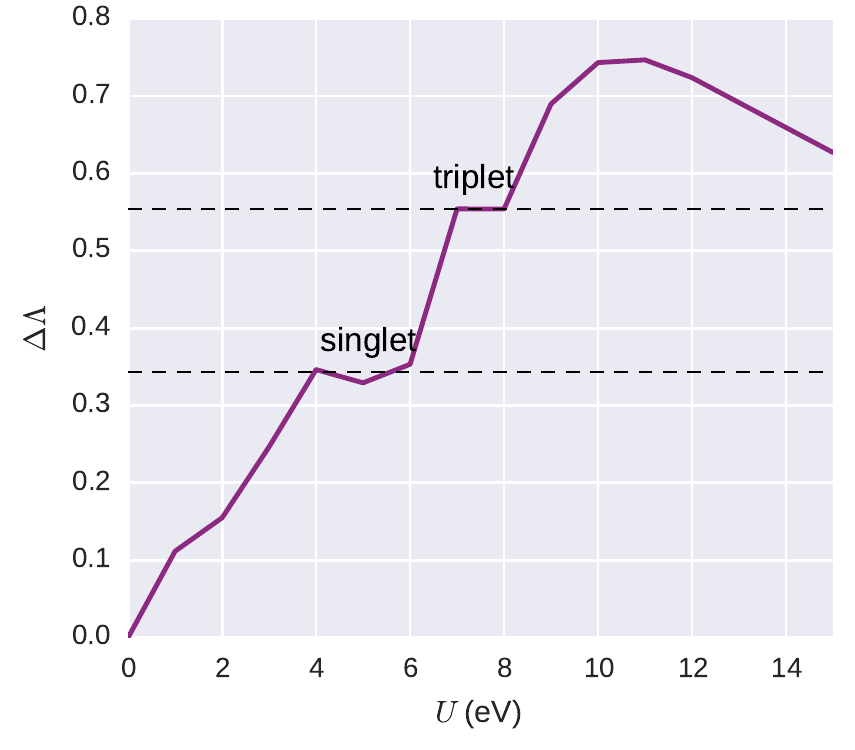}
\caption{The Von Neumann entropy $\Lambda$ of the reduced density matrix and its dependence on the on-site interaction $U$.}
\label{fig_entropy}
\end{center}
\end{figure}

The importance of multi-determinental physics can be quantified by the Von Neumann entropy. This provides a measure to identify the degree to which the system is in a mixed quantum state 
with a multitude of significant components. The Von Neumann entropy, obtained
in the di-Cu $3d$ subspace, is given by $\Lambda = \Tr \left[\hat{\rho}_d \log \hat{\rho}_d \right] $,
where $\hat{\rho}_d$ is the di-Cu reduced finite-temperature density matrix, traced over the
states of the AIM bath environment (Fig.~\ref{fig_entropy}). Interestingly, we find that the
entanglement entropy increases monotonically in the range
$U=0-10$\,eV. We note the presence of two plateaus,
for $U=4-6$\,eV and $U=7-8$\,eV, that coincide with the formation
of the singlet and triplet configurations in the histogram in Fig.~3 of the main text.

\newpage
\section*{Natural bond orbital analysis} 
In order to understand the nature of the bonding in the \butterfly\
complex, natural bond orbital (NBO) analysis was performed on the 
DFT and DMFT electronic densities.\cite{Reed1988a,nbo5,Lee2013a} This 
involves a series of diagonalisation and occupancy-weighted 
orthogonalisation procedures on the single-particle density 
matrix, transforming it into a set of atom-centered orthogonal 
natural atomic orbitals (NAOs), then natural hybrid orbitals, 
and finally the natural bond orbitals $\{\ket{\sigma_i}\}$, 
which are either one- or two-atom centered. 
By construction, this procedure decomposes the electronic 
density into terms resembling Lewis-type chemistry (with bonding and lone pairs of electrons).
The NBOs generated from DFT\,+\,DMFT densities largely retain the familiar profile 
of DFT-based NBOs, but their occupancies may be 
expected to deviate further from integer values due to quantum-mechanical
and finite-temperature multi-reference effects captured within DFT\,+\,DMFT.

\begin{figure}[h!]
\begin{center}
\includegraphics[width=0.8\columnwidth]{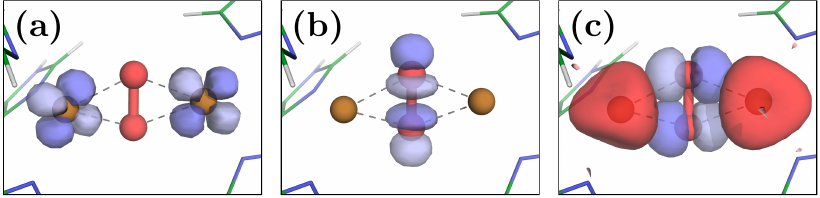}
\caption{Isosurfaces of several natural bonding orbitals for $U = 8$\,eV. (a) Two Cu $3d$ orbitals are identified as half-filled by the NBO analysis. (b) The O\textsubscript{2} $\sigma^*$ anti-bond is empty, and does not hybridise with any Cu orbitals. (c) Instead, O $2p$ (blue) to Cu $4s$ (red) charge transfer is favourable.}
\label{fig_isosurfaces}
\end{center}
\end{figure}

For hemocyanin, this analysis reveals a hole in one $3d$ orbital 
for each Cu atom (with $3d$ occupancies of 9.11 and 9.07 for $U=8$\,eV), 
confirming the expected Cu(II) oxidation state 3$d^{9}$4$s^{0}$ (Table~\ref{table_3d_occupations} and Figure~\ref{fig_isosurfaces}(a)).

\begin{table}[h!]
  \centering 
  \caption{The DMFT $3d$ orbital occupations of Cu in our model of ligated hemocyanin for different Hubbard $U$ values, as calculated using NBO. Note that the orbital labels correspond to the local axes to each Cu atom (Fig.~\ref{fig_local_axes}). All eight other Cu $3d$ orbitals had occupancies $> 1.98$ for all values of $U$.}
\begin{tabular}{c c c c c}
 \\ \hline
 \hline
 & \multicolumn{4}{c}{$U$ (eV)} \\
 \cline{2-5}
 & 0 & 4 & 6 & 8 \\
 \hline
 Cu\textsubscript{A} $d_{xz}$ & 1.55 & 1.47 & 1.27 & 1.17 \\
 Cu\textsubscript{B} $d_{xy}$ & 1.59 & 1.51 & 1.22 & 1.13 \\
  \hline
  \hline
\end{tabular}
\label{table_3d_occupations}
\end{table} 

\vspace{1cm}

A second-order perturbation analysis detects
multiple energetically favourable transfers of electronic density 
from filled to unfilled NBOs, revealing those aspects of the
 electronic structure that are not well described by Lewis-like chemistry. 
Early studies of hemocyanin identified back-bonding charge transfer 
from Cu $3d$ to oxygen $\sigma^*$ anti-bonding orbitals 
(Figure \ref{fig_isosurfaces}(b)) as an important factor in 
explaining the comparatively low 750\,cm\textsuperscript{-1} Raman 
frequency of the O\textsubscript{2} bond.\cite{Baldwin1992} However, 
our second-order perturbation analyses find that this back-transfer
is not present. For $U = 8$\,eV we instead detect 
favourable charge transfer from O $2p$ orbitals to Cu $4s$ orbitals (Figure \ref{fig_isosurfaces}(c)).

\clearpage


\newpage

\section*{Density of states}
\begin{figure}[!ht]
\begin{center}
\includegraphics[width=0.65\columnwidth]{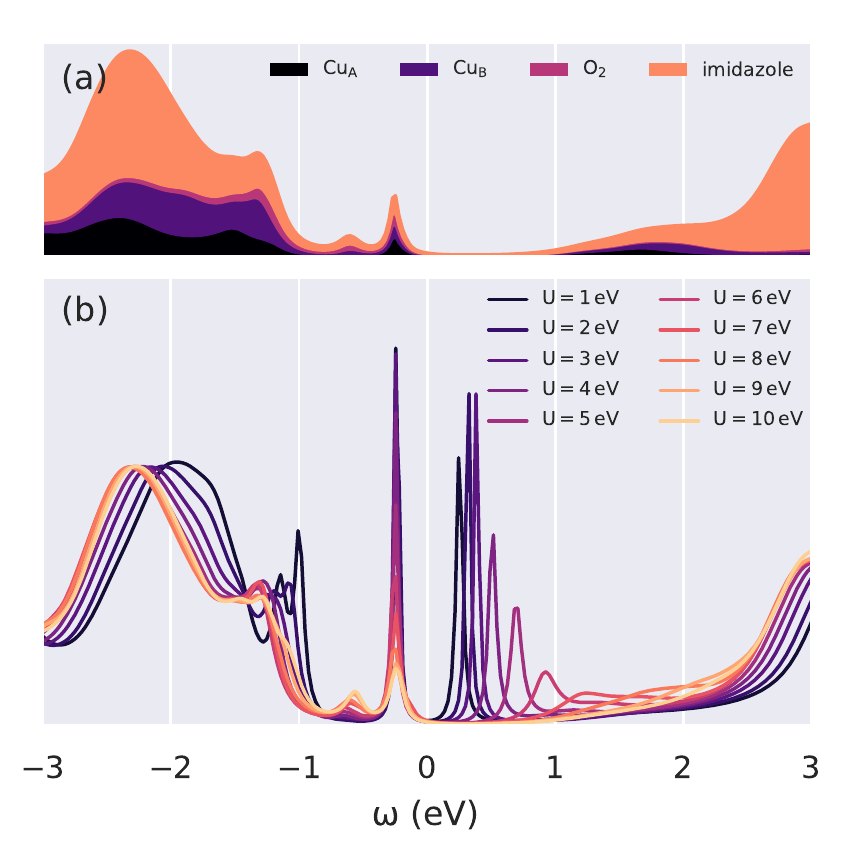}
\caption{(a) The local density of states for $U$ = 8\,eV and (b) the different total density of states of the system for a range of Hubbard $U$ values from DMFT.}
\label{fig_total_dos}
\end{center}
\end{figure}

Local and total densities of states for the \butterfly\ functional complex are shown in Fig.~\ref{fig_total_dos}.

\newpage
\section*{The atomic coordinates of the model}

The atomic coordinates (in Angstroms) of the model system are listed below and published separately in an xyz file. This is the QM region of the QM/MM region of Ref.~\onlinecite{Saito2014} optimised using the B3LYP hybrid functional. It closely matches the geometry of the \butterfly\ core in XRD measurements of oxyHc and oxyTy.

\ttfamily
\footnotesize
\begin{longtable}[l]{l r r r}
N  & 16.51891392 &  15.23078843  &  9.22976689 \\
H  & 16.93119882 &  15.08539696  &  8.30438809 \\
C  & 16.94902112 &  16.13114758  & 10.17383386 \\
N  & 15.18637912 &  14.99749680  & 10.95580106 \\
C  & 16.10933852 &  15.98208447  & 11.24036986 \\
H  & 16.10095092 &  16.50742455  & 12.17967036 \\
C  & 15.45961072 &  14.56963464  &  9.73009877 \\
H  & 14.93592812 &  13.80224391  &  9.18792729 \\
N  & 15.68221352 &  11.68923875  & 15.07079896 \\
H  & 15.62223022 &  10.91935273  & 15.75177196 \\
C  & 16.82912232 &  12.31709786  & 14.63750076 \\
N  & 15.03728172 &  13.21411391  & 13.63477906 \\
C  & 16.41234182 &  13.26768084  & 13.74798876 \\
H  & 16.99541732 &  13.97028085  & 13.18028166 \\
C  & 14.63155912 &  12.24369975  & 14.44250686 \\
H  & 13.62176392 &  11.89057990  & 14.57154846 \\
N  & 11.20316861 &  16.38796731  & 14.67452056 \\
H  & 10.83895228 &  16.68400689  & 15.58698756 \\
C  & 10.65158961 &  16.72090836  & 13.45946316 \\
N  & 12.41187546 &  15.35470124  & 13.15331656 \\
C  & 11.41667864 &  16.07778569  & 12.52477076 \\
H  & 11.32106945 &  16.09724541  & 11.45289136 \\
C  & 12.25019047 &  15.56909772  & 14.44942096 \\
H  & 12.86310074 &  15.16461678  & 15.23767766 \\
N  &  9.49463341 &   9.86374493  & 13.93094066 \\
H  &  9.73159525 &   9.00103207  & 14.43246936 \\
C  &  8.24884198 &  10.31835957  & 13.54789546 \\
N  &  9.84466860 &  11.48322602  & 12.50309476 \\
C  &  8.48328564 &  11.32677766  & 12.65632446 \\
H  &  7.78488610 &  11.93928595  & 12.11399846 \\
C  & 10.42165444 &  10.58129044  & 13.27870086 \\
H  & 11.47834322 &  10.42322846  & 13.40566166 \\
N  & 12.01234028 &   9.47930928  &  8.02845959 \\
H  & 12.68808052 &   9.09180375  &  7.36268858 \\
C  & 11.00609562 &   8.75784417  &  8.62913327 \\
N  & 11.09392331 &  10.81562920  &  9.51444153 \\
C  & 10.43873522 &   9.60240193  &  9.54022602 \\
H  &  9.61370823 &   9.41996499  & 10.20636756 \\
C  & 12.04142449 &  10.69948407  &  8.60142712 \\
H  & 12.77187130 &  11.45055695  &  8.35522084 \\
N  &  8.61465523 &  14.51152774  &  8.41694931 \\
H  &  8.38643811 &  14.76536198  &  7.45209746 \\
C  &  8.18880818 &  15.20575448  &  9.51961235 \\
N  &  9.53424857 &  13.53805896  & 10.13587576 \\
C  &  8.76071236 &  14.58675404  & 10.59130386 \\
H  &  8.67891762 &  14.82165976  & 11.63805866 \\
C  &  9.42431137 &  13.52365895  &  8.81797604 \\
H  &  9.92450163 &  12.84226676  &  8.15381340 \\
Cu &  10.90108130&   12.43349095 &  11.08418656 \\
Cu &  13.85734532&   14.03834172 &  12.18094446 \\
O  & 12.72653935 &  12.32631179  & 11.81473546 \\
O  & 12.74707482 &  13.23837395  & 10.72427086 \\
H  & 17.81251122 &  16.77851220  & 10.02095606 \\
H  & 17.81738922 &  12.01153775  & 14.98105336 \\
H  &  9.84447262 &  17.44637512  & 13.35784536 \\
H  &  7.32911179 &   9.83963986  & 13.88403226 \\
H  & 10.86685059 &   7.70012590  &  8.40572647 \\
H  &  7.51362297 &  16.05692406  &  9.43184952 \\
\end{longtable}
\rmfamily
\normalsize

\newpage
\section*{Convergence of the system-to-AIM mapping}
\begin{figure}[h!]
\centering
\includegraphics[width=0.55\columnwidth]{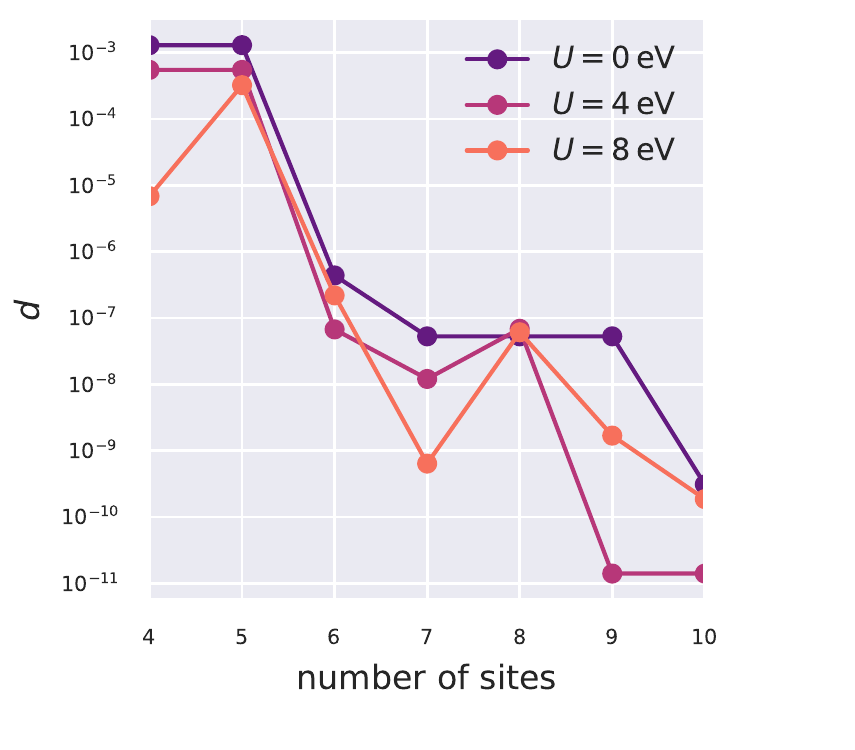}
\caption{The convergence of the distance $d$ as a function of the total number of sites (impurity and bath).}
\label{fig_convergence}
\end{figure}

In DMFT, the accurate mapping of the physical system to an Anderson Impurity Model (AIM) is crucial. This mapping is achieved by fitting the AIM hybridisation function $\Delta_\textrm{imp}(\omega)$ to the hybridisation function of the physical system $\tilde \Delta(\omega)$, where these hybridisation functions are defined as $\Delta(\omega) = \omega + \mu - t - G^{-1}(\omega) - \Sigma(\omega)$ (where $\mu$ is the chemical potential, $t$ the $d$-$d$ hopping parameters, $G$ the one-particle Green's function and $\Sigma$ the self-energy). In the impurity Green's function of an AIM, all of the parameters associated with the bath orbitals only appear via the hybridisation function; that is, it contains all of the physics of the bath orbitals pertinent to the impurity orbitals. The matching of the hybridisation functions is achieved by minimising the function
\begin{equation}
d = \sum_{\omega < \omega_c}\frac{1}{\omega^\gamma} \left| \Delta_\text{imp}(\omega) - \tilde \Delta(\omega) \right|^2
\label{eqn:DMFT_distance_function}
\end{equation}
with respect to the AIM parameters. $\omega_c$ is a cutoff frequency and $\gamma$ is a user-specified parameter that can allow for preferential weighting of agreement at low frequencies. If we increase the number of sites of our AIM, the AIM Hamiltonian has more parameters, and we stand a better chance of fitting the physical hybridisation function because the AIM hybridisation function is more flexible. Fig.~\ref{fig_convergence} shows the convergence of the distance $d$ as a function of the total number of sites in the AIM. We note that as a rule of thumb $d < 10^{-7}$ is generally adequate; if $d$ is much smaller than this it tends to indicate overfitting. The calculations in this work used eight sites.

\newpage

\newpage

\section*{Local axes}
\begin{figure}[h!]
\begin{center}
\includegraphics[width=0.6\columnwidth]{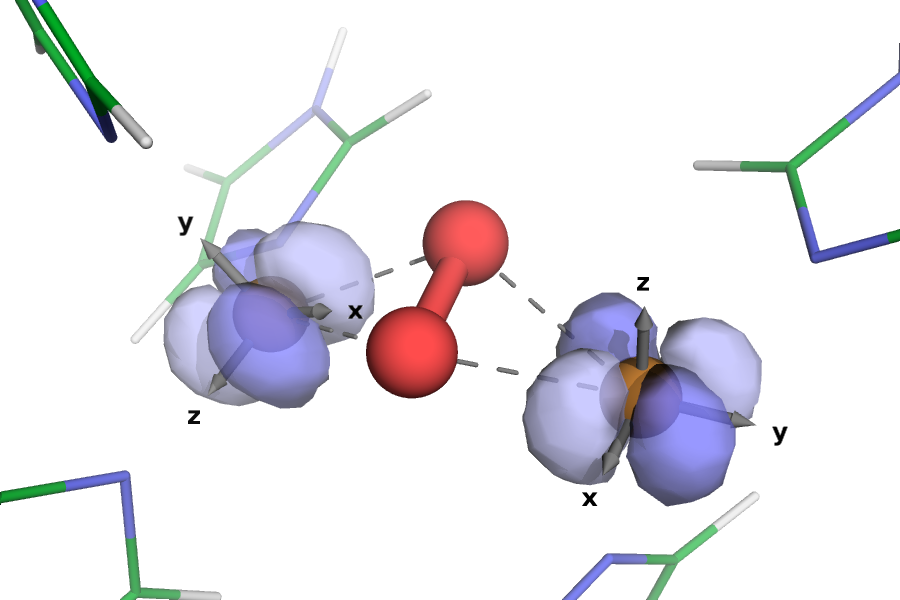}
\caption{The local axes for the Cu $3d$ correlated subspaces, and the two half-filled NBOs for comparison.}
\label{fig_local_axes}
\end{center}
\end{figure}

The local axes for the two Cu subspaces are shown in Fig.~\ref{fig_local_axes}. This particular rotation minimises the off-diagonal elements of the local Green's function. As shown in Table  \ref{table_3d_occupations} these axes localise the holes on single $d$ orbitals; $d_{xz}$ for Cu\textsubscript{A} (on the left of the figure) and $d_{xy}$ for Cu\textsubscript{B} (on the right). The two NBOs identified as being half-filled are plotted for comparison. Since the NBO analysis is agnostic to the projection procedure used in the cluster DMFT calculations, it is reassuring that these orbitals align with the axes.
 
\newpage
\section*{References}
\bibliography{biblio_updated}